%% file: main.tex
\documentclass[11pt]{article}
\input{preamble}

\title{Geography in Online Capital Allocation: Evidence from Equity-Based Crowdfunding}

\author{
Keiichi KAWAI\thanks{Keio University, Faculty of Economics, Keio University, 2-15-45 Mita, Minato-ku, Tokyo 108-8345 Japan (email: \href{mailto:keiichikawai@keio.jp}{\url{keiichikawai@keio.jp}}).}
\and %
Akira MATSUSHITA\thanks{University of Tokyo, Department of Economics, University of Tokyo, 7-3-1 Hongo, Bunkyo-ku, Tokyo 113-8654 Japan (email: \href{mailto:amatsushita@e.u-tokyo.ac.jp}{\url{amatsushita@e.u-tokyo.ac.jp}}).}
\thanks{The first author gratefully acknowledges the financial support from Keio University, the Zengin Foundation for Studies on Economics and Finance, Nomura Management School, The Tokio Marine Kagami Memorial Foundation, and the JSPS KAKENHI Grant (23K01313). The second author gratefully acknowledges the financial support from JSPS KAKENHI Grant (21J11717). We also extend our sincere gratitude to Fundinno for generously providing the data used in this analysis. We confirm that we have no conflicts of interest related to this data provision and have not received any financial support or incentives from Fundinno in connection with this research. An earlier version of this paper circulated under the title ``Local Support or Structural Barriers? Unpacking Local Bias in Equity-Based Crowdfunding.''
}
}

\date{July 2026}

\begin{document}
\maketitle
\input{abstract}
\input{sec_01_introduction}
\input{sec_02_institutional_background}
\input{sec_03_data_variables}

\input{sec_04_empirical_strategy}
\input{sec_05_results}
\FloatBarrier
\input{conclusion}

\bibliographystyle{aer}
\bibliography{ecf}
\newpage
\input{appendix}
\end{document}

%% file: preamble.tex
\usepackage[T1]{fontenc}
\usepackage[utf8]{inputenc}
\usepackage{mathpazo}
\usepackage{natbib}
\usepackage{amsmath,amsfonts,amssymb}

\usepackage{booktabs}
\usepackage{longtable}
\usepackage{array}
\usepackage{dcolumn}
\usepackage{threeparttable}
\usepackage{float}
\usepackage{makecell}
\usepackage{placeins}
\usepackage{caption}

\usepackage[hidelinks]{hyperref}
\usepackage{cleveref}
\hypersetup{
  pdftitle={Geography in Online Capital Allocation: Evidence from Equity-Based Crowdfunding},
  pdfauthor={Keiichi Kawai and Akira Matsushita}
}

\newcolumntype{d}[1]{D{.}{.}{#1}}

%% file: abstract.tex
\begin{abstract}

Digital investment platforms reduce search costs, yet realized investments can
remain geographically concentrated. Such concentration alone cannot distinguish
whether local investment reflects initial salience, proximity-related
information, prior ties, or nonpecuniary motives such as support for firms in
one's home prefecture. We use user-campaign records from Fundinno, Japan's
dominant equity-based crowdfunding platform, which link listing exposure,
campaign-page views, and investments to user residence and issuer location. For campaigns
outside the Tokyo metropolitan area (non-TMA), same-prefecture users are 6.7
percentage points more likely to invest after viewing a campaign page, relative
to an other-prefecture investment rate of roughly 10 percent. The corresponding
adjacent-prefecture difference is only 0.9 percentage points, and the
same-prefecture premium is small for campaigns inside the Tokyo metropolitan
area. The premium therefore remains after observed campaign-page access, is much
larger than the adjacent-prefecture difference, and is concentrated outside the
metropolitan core. These findings show that online access does not make
investment demand geographically neutral: for regional issuers,
same-prefecture investors remain disproportionately important even after
campaign-page access. The data do not identify the underlying motive, but the
pattern is difficult to explain by discovery alone or by smooth proximity.
\end{abstract}

%% file: sec_01_introduction.tex
\section{Introduction}
\label{sec:intro}

Digital financial platforms reduce search costs and present investment
opportunities through a common online interface. These features can weaken
the role of geography in capital markets by connecting firms and investors
across regions. Yet geographic concentration in online finance is difficult to
interpret from realized investments alone. A local investment premium could
arise before campaign-page access if local or nearby campaigns are more
salient. It could also arise after campaign-page access if same-prefecture or
nearby investors have proximity-related information, prior ties, or
nonpecuniary motives such as support for firms in their home prefecture.
Distinguishing among these explanations matters for whether online platforms
mainly reduce search frictions or also leave local capital-supply channels in
place.

To narrow the interpretation, we examine an online setting where
both user choices and geography are observable at each stage of the investment
process. Equity-based crowdfunding (ECF) fits this requirement because it
combines securities investment with platform-recorded intermediate choices. In
ECF, unlike donation- or reward-based crowdfunding, investors acquire claims on
unlisted firms and have financial
stakes in campaign outcomes. Because these issuers are small unlisted firms,
investors are less likely to have publicly available information about them
outside the platform, making platform-provided disclosure a central part of the
information environment. Campaigns are presented through a common platform
interface, so the campaign information supplied by the platform is largely
standardized across investors. ECF platforms also record the path from listing
exposure to campaign-page access and investment. These records allow us to
analyze local concentration through intermediate choices, not only through
final funding outcomes.

We use proprietary user-campaign records from Fundinno, Japan's dominant
equity-based crowdfunding platform. The data link listing-page exposure,
campaign detail-page views, investments, user residence, and campaign firm
location. In the raw data, investment is locally concentrated, with users more
likely to invest in campaigns from their own prefecture. The empirical question
is not whether local concentration exists, but which interpretation best fits
the observed sequence of exposure, campaign-page access, investment, and
user-campaign geography.

We use two conditional margins of the online choice process to separate initial
inspection from investment after campaign-page access. Detail-page viewing
conditional on listing-page exposure
measures whether an exposed user chooses to inspect a campaign. Positive net
investment conditional on campaign-page viewing measures whether a user invests
after accessing detailed campaign information. A same-prefecture difference at
the first margin is naturally related to initial attention or discovery,
whereas a difference at the second margin remains after campaign-page access.
We estimate both probabilities with user and campaign fixed effects. Campaign
fixed effects compare users facing the same campaign, while user fixed effects
compare a given user's choices across campaigns.

A same-prefecture premium in the view-to-investment margin does not by itself
identify the underlying mechanism. We therefore combine the two conditional
margins with two geographic comparisons. The adjacent-prefecture
comparison asks whether the premium extends to nearby but nonlocal users. The
TMA/non-TMA comparison asks whether the premium is concentrated outside the
metropolitan core, where same-prefecture status is more likely to correspond
to a distinct local context. Inside the TMA, prefectural boundaries often cut
across commuting patterns and offline networks. The TMA/non-TMA split also
matters because campaign supply is heavily concentrated in the TMA, so pooled
estimates would be dominated by metropolitan campaigns. It does not identify
the underlying motive: the remaining premium may still reflect local
information, prior ties, issuer outreach, or home-prefecture support motives,
including local attachment.

The estimates reveal a same-prefecture premium concentrated outside the
metropolitan core. In the view-to-investment margin, same-prefecture users are
6.7 percentage points more likely to invest in non-TMA campaigns after viewing
a campaign page, relative to an other-prefecture investment rate of roughly 10
percent. The corresponding adjacent-prefecture difference is 0.9 percentage
points, and the same-prefecture differences for TMA campaigns are small in
economic magnitude. The key finding is therefore a remaining same-prefecture
premium: it appears after observed campaign-page access, is much larger than
the adjacent-prefecture difference, and is concentrated outside the metropolitan
core. Because the premium remains after users access the campaign page, a pure
discovery account is incomplete. Because the adjacent-prefecture estimate
is much smaller, a smooth proximity account fits the data less well. The ECF
setting also matters: issuers are small unlisted firms, and campaign
information is supplied through a common platform page, so publicly available
information is less likely to be the whole explanation. The results do not
identify the underlying motive, but they leave home-prefecture support as a
possible contributor alongside local information, prior ties, and issuer
outreach.

These findings imply that online access and geographic neutrality are distinct
concepts. For regional issuers, especially outside the metropolitan core,
same-prefecture users may remain a meaningful source of investment demand after
campaign-page access. Online equity finance can expand access without making
investment demand geographically neutral.

This paper contributes to the literature on geography in online investment
markets. A central finding in this literature is that online platforms do not
eliminate geography: platform participants remain more likely to finance
nearby projects even when opportunities are presented online
\citep[e.g.,][]{Agrawal2015-fr,Lin2016-kw,Guenther,Bade2021-uq,Hornuf2022-fs,Cai2024-eg}.
Prior work has attributed such patterns to information, networks, or investor
preferences. Because much prior work observes only realized investments or
aggregate funding outcomes, these channels are difficult to disentangle.

Our contribution is to show that the relevant geography is not only a
pre-access attention margin. Listing exposure and campaign-page views allow us
to separate initial inspection from investment after campaign-page access. The
same-prefecture premium persists at this investment margin, is much larger than
the adjacent-prefecture difference, and is concentrated outside the
metropolitan core. The paper also adds to the emerging evidence on Japanese ECF
\citep{Honjo2023-tl,Nose2023-bk} by shifting attention from campaign-level
success and signaling to investor-side geographic allocation.

The rest of the paper is organized as follows. \Cref{sec:institution} describes
the Japanese ECF setting and the Fundinno platform. \Cref{sec:data_variables}
introduces the data, sample construction, and key variables.
\Cref{sec:empirical_strategy} describes the empirical framework.
\Cref{sec:results} presents the results.

%% file: sec_02_institutional_background.tex
\section{Institutional Background and Research Setting}
\label{sec:institution}

This section explains why Japanese equity-based crowdfunding (ECF) on Fundinno
provides a useful setting for interpreting local concentration in online
capital allocation.
ECF is a regulated securities market involving small unlisted issuers. Because
these firms are unlikely to be widely known, platform disclosure is central to
the information environment facing most investors, even though local
information and prior familiarity may still vary across user-campaign pairs.
Fundinno records
the online choice process from listing-page exposure to detail-page viewing and
investment, and the data identify both user residence and issuer location at
the prefecture level. Regulation and campaign terms make campaign-level
participation the natural empirical margin. Finally, because Fundinno accounts
for most Japanese ECF activity during the sample period, we observe
user-campaign activity within a single platform and cover most of the market.

\subsection{Regulated Securities Market and Platform Disclosure}

ECF in Japan was introduced through amendments to the Financial Instruments and Exchange Act (FIEA) and related subordinate regulations that were implemented in 2015.\footnote{The relevant legal framework includes the Financial Instruments and Exchange Act, Act No. 25 of 1948; the Order for Enforcement of the Financial Instruments and Exchange Act, Cabinet Order No. 321 of 1965; and the Cabinet Office Order on Financial Instruments Business, Cabinet Office Order No. 52 of 2007.} For institutional overviews of the Japanese ECF market, see \citet{Kato2021-yr} and \citet{Inaba2023-cv}. The Japanese framework treats ECF as a regulated securities business rather than as a lightly regulated fundraising channel. In the terminology of Japanese securities regulation, ECF offerings are regulated as small-amount electronic solicitation for unlisted securities. The framework allows startups and small firms to raise equity capital from retail investors through online platforms, subject to investor-protection rules.

Platforms intermediate these offerings and present campaign information in a common online format, including business plans, offering terms, risk information, and financial information. Solicitation takes place through the platform website and other electronic communications rather than through conventional face-to-face securities sales. This platform-based disclosure structure matters because investors evaluate campaigns through the same interface. The platform supplies a common set of campaign materials and records the sequence from initial exposure to investment.

Because ECF issuers are small unlisted firms, and the campaigns in our sample
typically involve modest fundraising amounts
(\Cref{tab:campaign_characteristics}), investors are unlikely to have the kind
of publicly available information that exists for listed or widely known firms.
Platform disclosure therefore represents a central part of the information
environment for most users, even though local information and prior familiarity
may vary by user-campaign pair. This does not make information identical across
investors, but it makes observed campaign-page access a meaningful conditioning
event: after this point, all viewers have access to the same platform-provided
disclosure.

\subsection{Market Structure and the Role of Fundinno}

Fundinno is the dominant platform in the Japanese ECF market during our sample
period. In campaign-level records from the Japan Securities Dealers Association
(JSDA), Fundinno accounts for 569 of 715 Japanese ECF offerings with
subscription periods ending by October 31, 2024, or 79.6\% of offerings. By
total issue value, Fundinno accounts for \yen12.54 billion out of \yen15.33
billion, or 81.8\% of the market
\citep{JSDA2026ECFStatistics}.\footnote{The calculation includes both stock and share-subscription-right offerings. Fundinno includes offerings listed under its former name, Japan Cloud Capital, and under its current name. \Cref{sec:data_variables} describes the corresponding proprietary campaign and user-campaign analysis samples.}

Fundinno's dominance allows us to observe most Japanese ECF activity within a
common platform environment. Geographic differences are therefore less likely
to reflect cross-platform variation in platform design, issuer screening, or
investor composition.

\subsection{Fundinno Campaign Timeline}

Fundinno campaigns follow a sequence that corresponds closely to our empirical design. Each campaign typically begins with a preview phase lasting around two weeks, during which users may view the campaign page and read detailed campaign information but cannot yet invest. This is followed by an investment phase, which is generally much shorter, often lasting only a few days. During this phase, users may invest until the campaign either reaches its funding limit or the offering period ends. Campaigns follow an all-or-nothing rule, so funds are transferred only if the minimum funding threshold is met.

This phased structure is important because users can access detailed campaign
information before making an investment application. It therefore supports the
distinction between campaign-page access and subsequent investment.
The platform navigation structure defines the listing-page exposure margin.

\subsection{Platform Navigation and Initial Exposure}

Users may arrive at a campaign's detail page through two main routes. The first is through the platform's listing page, where campaigns are presented in a grid format. Each campaign listing contains a thumbnail image, a short description, information on progress toward the funding target, and the remaining time. From the listing page, users may click on a listing to access the corresponding campaign's detail page.

The second route is direct access. In this case, users arrive at the campaign page through external links, such as promotional emails or referral links, and therefore bypass the listing page. Regardless of the route, the campaign detail page provides much richer information, including the business plan, financial projections, and offering terms.

This distinction matters for our empirical analysis because the information available to the user differs across the two routes. Users click from the listing page based on limited and common-format information. After opening the detail page, they can access richer information before deciding whether to invest. The platform design therefore helps distinguish the attention margin from investment conditional on campaign-page access.

The two access routes correspond to different empirical margins. Listing-page exposure allows us to examine whether users proceed from a common campaign listing to a detail-page view. Direct access does not generate this listing-to-view transition, but once a detail-page view is observed, the subsequent investment decision can be analyzed in the same way. Because investment applications are made after users access the campaign detail page, we treat detail-page viewing as the conditioning event for the investment-stage analysis.

\subsection{Investment Limits and Campaign-Level Participation}

Two features of the Japanese ECF setting make campaign-level participation the
natural empirical margin. First, during our sample period, retail investors
subject to the ordinary cap were generally limited to investing at most
\yen500{,}000 in a single issuer through ECF, and issuer fundraising through
ECF was capped below \yen100 million within a one-year period.\footnote{These
limits were set under the FIEA framework and related subordinate regulations,
including the Order for Enforcement of the Financial Instruments and Exchange
Act and the Cabinet Office Order on Financial Instruments Business. They were
also reflected in the Japan Securities Dealers Association's Rules Concerning
Equity Crowdfunding Business.} These limits keep offerings relatively small and
limit the influence of a small number of large investors on campaign outcomes.

Second, campaign terms make the investment decision discrete. Each campaign
specifies a minimum investment amount, which is often around \yen100{,}000.
Together with the per-issuer investment cap, this limits the range of feasible
investment amounts for ordinary retail investors. We therefore model the
relevant decision as whether a user participates in a campaign, rather than as a
continuous portfolio-allocation choice over investment amounts.

These features distinguish the setting from environments in which investors
allocate a flexible budget across many small positions. In the Japanese ECF
market, we model investor behavior as a sequence of campaign-specific choices:
whether to access a campaign detail page and, conditional on doing so, whether
to participate by investing. We therefore define the investment-stage outcome as
positive net investment at the user-campaign level. This extensive-margin
outcome captures whether a user supplies capital to a campaign after
campaign-page access.

The regulatory limits described above are the relevant constraints for our
sample period. Reforms associated with the specified-investor system allowed
eligible specified investors to invest above the ordinary per-issuer cap during
our sample period \citep{FUNDINNO2022SpecifiedInvestor}. The ordinary investor
limit was raised only after the end of our sample period; for a current
institutional description of the post-sample investor limit, see
\citet{JSDA2026ECFOverview}. Our analysis covers platform activity through
October 31, 2024, so the \yen500{,}000 ordinary-investor cap is the relevant
baseline for the main sample.

\subsection{Geographic Information and the Role of Prefecture-Level Location}

Campaign and investor locations are observed at the prefecture level. This
allows us to identify user-campaign pairs in the same prefecture and to compare
them with adjacent-prefecture pairs, which is central to distinguishing broad
proximity from same-prefecture locality. The same location information also
allows us to classify campaigns by whether they are located inside or outside
the Tokyo metropolitan area.

Some campaign listings also display geographic information, such as a
prefecture or city name, in the title or short description. Because this
information appears before the user opens the detail page, the display
indicator used below captures listing-stage geographic information. These
features define the two observed margins used below and record whether
geographic information appears before detail-page access. The next section
describes how we construct the user-campaign samples and variables.

%% file: sec_03_data_variables.tex
\section{Data and Variable Construction}
\label{sec:data_variables}

This section describes how the Fundinno records are used to construct the
user-campaign data that link the online choice sequence to user-campaign
geography.
The key feature of the data is that platform exposure, campaign-page viewing,
investment, user residence, and issuer location are linked at the user-campaign
level. We first describe the
administrative data sources and the construction of the campaign and
user-campaign samples. We then summarize the sample's scale and geographic
composition, define the key variables, and report descriptive transition rates
for the two observed margins.

\subsection{Data Sources}

Our analysis uses proprietary administrative data provided by Fundinno, covering
campaigns and user-campaign activity on the platform through October 31, 2024.
We observe not only whether a user invested in a campaign, but also whether the
same user was exposed to the campaign on the platform and whether the user
viewed the campaign detail page before investing.

\Cref{tab:data_sources} summarizes the main administrative data sources used to
construct the user-campaign records. The campaign records contain campaign
characteristics, schedules, offering terms, and issuer location. The user
records contain user demographics and residence prefecture. Our
user-level analysis is restricted to users who have completed the required
identity and eligibility checks needed to invest on the platform. The order
records contain raw investment applications, cancellations, and waitlist-related
events. These events are also used to define the main positive net-investment
indicator. The access panel summarizes user-campaign platform activity and is
used to distinguish listing-page exposure from campaign detail-page viewing.

\Cref{tab:data_sources} also reports the number of records in each main data
source. The access panel is much larger than the other data sources because it
records user-campaign activity at the access-log level; the analysis samples
constructed from this panel are reported below.

\begin{table}[htbp]
  \centering
  \caption{Overview of Administrative Datasets}
  \label{tab:data_sources}
  \begin{tabular}{@{}p{3.2cm}p{11.0cm}@{}}
    \toprule
    \textbf{Dataset} & \textbf{Key Information Provided} \\
    \midrule
    Campaign records\newline
    (569 rows) & Campaign schedules, funding targets and limits, offering terms, firm characteristics, campaign outcome, headquarters prefecture, and whether the listing text displays geographic information. \\
    \addlinespace
    User records\newline
    (48,669 rows) & User residence prefecture, registration timing, identity-check and eligibility status, demographic characteristics, income and asset categories, and prior investment experience. \\
    \addlinespace
    Order records\newline
    (119,443 rows) & Raw investment applications, cancellations, and waitlist-related order events, including timestamps and application values. These events are used to define positive net investment after cancellations and unpaid orders are removed. \\
    \addlinespace
    Access panel\newline
    (42,804,752 rows) & User-campaign access records used to identify listing-page exposure, campaign detail-page views, and the timing of those events relative to investment. \\
    \bottomrule
\end{tabular}
\end{table}
\FloatBarrier

\subsection{Analysis Sample Construction}

The Fundinno campaign records contain 569 offerings with subscription periods
ending by October 31, 2024, matching the public count reported in
\Cref{sec:institution}. Some of these offerings cannot be used in the
user-campaign analysis because the access-log or investment records needed to
construct comparable margin outcomes are missing or internally inconsistent.
The exclusions are small relative to platform coverage: the final campaign
sample contains 527 campaigns.
We make these exclusions in two steps.

We first exclude 30 campaigns before constructing the user-campaign panel: 26
early campaigns without reliable user-level access logs and four campaigns
that were canceled, stopped, withdrawn, or classified in the source data as
post-campaign failures, meaning that they were not treated as valid completed
offerings in the platform records. Of the remaining 539 campaigns, we exclude
12 additional campaigns from the
user-campaign analysis sample: six lack complete or internally consistent
investment and access-log records, and six lack usable access-panel coverage.
Thus, the analysis retains nearly all Fundinno offerings for which exposure,
viewing, investment, and campaign geography can be measured using the same
definitions.

We then construct the user-campaign analysis samples by merging campaign
characteristics, eligible-user information, net investment records, and access
records. The sample for the viewing margin contains 1,179,437 user-campaign
observations with a qualifying listing-page exposure while the campaign was
available on the platform. The sample for the investment margin contains
574,213 user-campaign observations in which the user viewed the campaign detail
page. These two samples define the observed denominators for the conditional
margins used in
the empirical analysis: listing-page exposure for the viewing margin and
campaign-page viewing for the investment margin. This organization keeps the
viewing analysis tied to observed listing exposure and the investment analysis
tied to observed campaign-page access. The corresponding outcome variables are
defined in \Cref{sec:key_variables}.

\subsection{Sample Characteristics}

Having defined the campaign and user-campaign analysis samples, we next
summarize the characteristics of the 527 campaigns used in the analysis.
\Cref{tab:campaign_characteristics} provides scale for the campaigns underlying
the user-campaign analysis by reporting campaign terms, platform activity, and
realized funding outcomes. The average campaign raised about \yen24.1 million
from 143 investors, consistent with the small-scale, unlisted-issuer setting
described in \Cref{sec:institution}.

\begin{table}[htbp]
\centering
\begin{threeparttable}
\caption{Campaign-Level Sample Characteristics}
\label{tab:campaign_characteristics}
\footnotesize
\begin{tabular}{@{}lrrrrr@{}}
    \toprule
    & \textbf{Mean} & \textbf{SD} & \textbf{25th pct.} & \textbf{Median} & \textbf{75th pct.} \\
    \midrule
    Funding target (million JPY) & 13.9 & 7.8 & 10.0 & 12.5 & 16.0 \\
    Funding limit (million JPY) & 48.4 & 21.8 & 30.1 & 45.0 & 60.0 \\
    Campaign duration (days) & 6.6 & 7.1 & 2.6 & 3.2 & 6.6 \\
    \midrule
    Listing-page exposures & 2,238 & 1,013 & 1,480 & 2,257 & 2,866 \\
    Detail-page viewers & 1,090 & 343 & 834 & 1,037 & 1,301 \\
    Investors per campaign & 143 & 118 & 50 & 111 & 199 \\
    Amount raised (million JPY) & 24.1 & 19.5 & 9.5 & 19.5 & 32.7 \\
    Success ratio (\%) & 194.3 & 155.1 & 79.8 & 150.0 & 274.2 \\
    \bottomrule
\end{tabular}
\begin{tablenotes}[flushleft]
\footnotesize
\item[] 
\textbf{Notes:} Entries are campaign-level summary statistics for the 527
campaigns in the user-campaign analysis sample. Funding variables are reported in
millions of yen. Success ratio is the amount raised divided by the funding
target, multiplied by 100. Listing-page exposures and detail-page viewers are
counts of user-campaign observations. Investors per campaign counts users with
positive net investment after cancellations and unpaid orders are removed.
\end{tablenotes}
\end{threeparttable}
\end{table}

Beyond scale and platform activity, \Cref{tab:geo_tma_distribution} documents
the geographic composition of the analysis sample. Campaign supply is
concentrated in the Tokyo metropolitan area (TMA: Tokyo, Kanagawa, Saitama, and
Chiba): 383 of the 527 campaigns (72.7\%) are located in the TMA, compared with
24,755 of 48,666 eligible users (50.9\%).
This imbalance motivates reporting TMA and non-TMA campaigns separately below.
Same-prefecture pairs may be related to different
forms of locality inside and outside the metropolitan core: in the TMA,
prefectural boundaries often cut across commuting and economic activity,
whereas outside the TMA they are more likely to correspond to a distinct local
setting. Given the concentration of campaign supply in the TMA,
the main analysis separates TMA and non-TMA campaigns rather than relying only
on pooled estimates. Because several non-TMA areas contain few campaigns and
few display listings, we treat the more disaggregated economic-area analysis as
supplementary.

\begin{table}[htbp]
\centering
\begin{threeparttable}
\caption{Geographic Distribution of the Analysis Sample}
\label{tab:geo_tma_distribution}
\footnotesize
\newcommand{\countpct}[2]{\makebox[3.3em][r]{#1}\hspace{0.2em}\makebox[3.9em][r]{(#2\%)}}
\begin{tabular}{@{}lccc@{}}
    \toprule
    Region & Campaigns & Display listings & Eligible users \\
    \midrule
    TMA & \countpct{383}{72.7} & \countpct{9}{2.3} & \countpct{24,755}{50.9} \\
    Non-TMA & \countpct{144}{27.3} & \countpct{24}{16.7} & \countpct{23,911}{49.1} \\
    \midrule
    Total & \countpct{527}{100.0} & \countpct{33}{6.3} & \countpct{48,666}{100.0} \\
    \bottomrule
\end{tabular}
\begin{tablenotes}[flushleft]
\footnotesize
\item[] 
\textbf{Notes:} Campaigns are classified by issuer headquarters; users by
residence. Percentages for campaigns and users are column shares; display
percentages are shares within region. Three users with missing residence
prefecture are excluded from user shares.
\end{tablenotes}
\end{threeparttable}
\end{table}
\FloatBarrier

\subsection{Key Variables and Descriptive Transition Rates}
\label{sec:key_variables}

We next define the user-campaign variables used to construct the two
conditional margins and the geographic comparisons in the empirical analysis.
\Cref{tab:notations_definitions} summarizes the notation and variable
definitions used throughout the empirical analysis. The main viewing-margin
outcome is \texttt{visit\_view}, which equals one when a qualifying listing-page
exposure leads to a campaign detail-page view. The main investment-margin
outcome is denoted by $\texttt{inv}_{i \to j}$. It equals one when the user has
a positive net investment in the campaign and the first observed detail-page
view occurs no later than the first positive net investment. The key geographic
variables are $\texttt{local}_{i \to j}$, which identifies same-prefecture
user-campaign pairs, and $\texttt{adj}_{i \to j}$, which identifies
adjacent-prefecture pairs excluding same-prefecture pairs. The variable
$\texttt{display}_{j}$ indicates whether the campaign listing displays a
prefecture or city name.

\begin{table}[htbp]
\centering
\begin{threeparttable}
\renewcommand{\arraystretch}{1.35}
\caption{Notation and Variable Definitions}
\label{tab:notations_definitions}
\footnotesize
\begin{tabular}{@{}p{3.0cm}p{11.6cm}@{}}
    \toprule
    \textbf{Variable} & \textbf{Definition / Explanation} \\
    \midrule
    $\texttt{visit}_{i \to j}$ &
    Equals 1 if user $i$ has a qualifying listing-page exposure to campaign $j$ while the campaign is available on the platform. \\
    \addlinespace
    $\texttt{view}_{i \to j}$ &
    Equals 1 if user $i$ views the detail page of campaign $j$. \\
    \addlinespace
    $\texttt{visit\_view}_{i \to j}$ &
    Main viewing-margin outcome. Defined for observations with $\texttt{visit}_{i \to j}=1$; equals 1 if the listing-page exposure leads to a campaign detail-page view. \\
    \addlinespace
    $\texttt{inv}_{i \to j}$ &
    Main investment-margin outcome. Defined for observations with $\texttt{view}_{i \to j}=1$; equals 1 if user $i$ has a positive net investment in campaign $j$ and the first observed detail-page view occurs no later than the first positive net investment. \\
    \addlinespace
    $\texttt{local}_{i \to j}$ &
    Equals 1 if user $i$'s residence prefecture is the same as the headquarters prefecture of the company raising funds through campaign $j$. \\
    \addlinespace
    $\texttt{adj}_{i \to j}$ &
    Equals 1 if the user's residence prefecture is adjacent to the campaign's company prefecture, excluding same-prefecture pairs. \\
    \addlinespace
    $\texttt{display}_{j}$ &
    Equals 1 if the campaign listing explicitly displays geographic information such as a prefecture or city name. \\
    \bottomrule
\end{tabular}

\begin{tablenotes}[flushleft]
\footnotesize
\item[] 
\textbf{Notes:} Same-prefecture pairs are coded as local rather than adjacent. The investment-margin outcome is based on positive net investment after cancellations and unpaid orders are removed.
\end{tablenotes}
\end{threeparttable}
\end{table}

Using these variables, we construct campaign-level descriptive rates for the two
observed user-campaign margins. For each campaign $j$ and user group $G$, the
two rates are
\[
\text{Listing-to-view rate}_{jG}
=
\frac{\sum_{i \in G_j} \texttt{visit\_view}_{i \to j}}
     {\sum_{i \in G_j} \texttt{visit}_{i \to j}},
\qquad
\text{View-to-investment rate}_{jG}
=
\frac{\sum_{i \in G_j} \texttt{inv}_{i \to j}}
     {\sum_{i \in G_j} \texttt{view}_{i \to j}}.
\]
\Cref{tab:analysis_sample_summary} reports the mean and standard deviation of
these campaign-level rates across campaigns.

Because the rates are first computed at the campaign level and then averaged
across campaigns, the table gives equal weight to each campaign rather than
weighting campaigns by the number of user-campaign observations. Panel A
reports the listing-to-view margin by whether geographic information is
displayed at the listing stage, separately for campaigns inside and outside the
Tokyo metropolitan area (TMA). Panel B reports the view-to-investment margin,
separately for TMA and non-TMA campaigns.

\begin{table}[htbp]
\centering
\begin{threeparttable}
\caption{Campaign-Level Descriptive Transition Rates}
\label{tab:analysis_sample_summary}
\footnotesize
\begin{tabular}{@{}p{4.2cm}r@{\ }l r@{\ }l r@{\ }l r@{\ }l@{}}
    \toprule
    & \multicolumn{4}{c}{\textbf{TMA campaigns}} &
    \multicolumn{4}{c}{\textbf{Non-TMA campaigns}} \\
    \cmidrule(lr){2-5}\cmidrule(l){6-9}
    & \multicolumn{2}{c}{\textbf{No display}} &
    \multicolumn{2}{c}{\textbf{Display}} &
    \multicolumn{2}{c}{\textbf{No display}} &
    \multicolumn{2}{c}{\textbf{Display}} \\
    & \multicolumn{2}{c}{\textbf{(374)}} &
    \multicolumn{2}{c}{\textbf{(9)}} &
    \multicolumn{2}{c}{\textbf{(120)}} &
    \multicolumn{2}{c}{\textbf{(24)}} \\
    \midrule
    \multicolumn{9}{@{}l}{\textbf{Panel A: Listing-to-view transition}} \\
    Same-prefecture users &
    27.9\% & (13.5) &
    22.8\% & (3.7) &
    29.3\% & (14.8) &
    39.5\% & (16.4) \\
    Other-prefecture users &
    27.8\% & (14.4) &
    21.4\% & (4.8) &
    23.9\% & (9.7) &
    20.9\% & (6.3) \\
    Adjacent-prefecture users &
    27.5\% & (13.9) &
    21.9\% & (5.0) &
    24.1\% & (12.6) &
    22.8\% & (9.4) \\
    \addlinespace
    \midrule
    & \multicolumn{4}{c}{\textbf{TMA campaigns}} &
    \multicolumn{4}{c}{\textbf{Non-TMA campaigns}} \\
    \cmidrule(lr){2-5}\cmidrule(l){6-9}
    & \multicolumn{4}{c}{\textbf{(383)}} &
    \multicolumn{4}{c}{\textbf{(144)}} \\
    \midrule
    \multicolumn{9}{@{}l}{\textbf{Panel B: View-to-investment transition}} \\
    Same-prefecture viewers &
    \multicolumn{4}{c}{14.5\% (9.6)} &
    \multicolumn{4}{c}{22.6\% (17.1)} \\
    Other-prefecture viewers &
    \multicolumn{4}{c}{12.6\% (9.6)} &
    \multicolumn{4}{c}{10.5\% (8.2)} \\
    Adjacent-prefecture viewers &
    \multicolumn{4}{c}{13.1\% (9.6)} &
    \multicolumn{4}{c}{11.2\% (10.5)} \\
    \bottomrule
\end{tabular}
\begin{tablenotes}[flushleft]
\footnotesize
\item[] 
\textbf{Notes:} Entries report campaign-level means; standard deviations,
in percentage points, are in parentheses. TMA denotes the Tokyo metropolitan area. Panel A
reports detail-page views per listing-page exposure and splits campaigns by
whether the listing displays geographic information. Panel B reports positive
net investments per detail-page view. Same-prefecture users are users whose
residence prefecture matches the campaign firm's headquarters prefecture.
Other-prefecture users exclude same-prefecture users. Adjacent-prefecture users
also exclude same-prefecture users; adjacent-prefecture means are calculated
over campaigns with adjacent-prefecture observations in the relevant stage. The
TMA-display column in Panel A is based on nine campaigns and should be
interpreted cautiously.
\end{tablenotes}
\end{threeparttable}
\end{table}

The descriptive rates preview the main geographic patterns examined below. For
non-TMA campaigns with geographic display, same-prefecture users have a
listing-to-view rate of 39.5 percent, compared with 20.9 percent for
other-prefecture users. In the view-to-investment margin for non-TMA campaigns,
same-prefecture viewers invest at a rate of 22.6 percent, compared with 10.5
percent for other-prefecture viewers and 11.2 percent for adjacent-prefecture
viewers. The corresponding TMA differences are much smaller: in the
view-to-investment margin, same-prefecture viewers invest at 14.5 percent,
compared with 12.6 percent for other-prefecture viewers and 13.1 percent for
adjacent-prefecture viewers.

%% file: sec_04_empirical_strategy.tex
\section{Empirical Framework and Strategy}
\label{sec:empirical_strategy}

This section moves from the campaign-level descriptive patterns in
\Cref{tab:analysis_sample_summary} to user-campaign fixed-effect comparisons.
The descriptive rates summarize the two observed margins--listing-to-view and
view-to-investment--by user-campaign geography. They suggest larger
same-prefecture differences for non-TMA campaigns and much smaller
adjacent-prefecture differences, but they are campaign-level averages and do
not control for persistent user differences or heterogeneity across ECF
campaigns. We therefore estimate conditional transition probabilities with user
and campaign fixed effects.

\subsection{Conditional Margins and Empirical Comparisons}

The empirical strategy relies on three related comparisons. The first asks
where the same-prefecture difference appears in the online choice sequence. A
difference in the listing-to-view margin captures initial attention or
inspection after listing-page exposure. A difference in the view-to-investment
margin captures investment after users have accessed the campaign detail page
and the richer information it contains.

The second comparison concerns nearby but nonlocal users. The
adjacent-prefecture indicator is a coarse proximity check rather than a
complete distance-gradient analysis. If the same-prefecture difference reflects
a smooth geographic proximity gradient, adjacent-prefecture viewers should also
exhibit higher investment probabilities. A difference concentrated in
same-prefecture pairs, with a much smaller adjacent-prefecture difference, is
therefore less consistent with a smooth proximity account, although it does not
rule out prefecture-level networks, prior familiarity, or regionally targeted
promotion.

The third comparison distinguishes TMA and non-TMA campaigns. Same-prefecture
status may have different interpretive content inside and outside the
metropolitan core, and campaign supply is concentrated in the TMA. Together,
these comparisons can weaken simple accounts based on discovery alone or on a
smooth proximity gradient, even though they do not identify a single motive. We
use same-prefecture locality as a descriptive label for the empirical pattern
in which the investment premium is concentrated among same-prefecture users
rather than nearby users more broadly, remains after campaign-page access, and
is concentrated outside the metropolitan core.

\subsection{Viewing-Margin Specification}

The viewing-margin sample consists of user-campaign observations with a
qualifying listing-page exposure, \(\texttt{visit}_{i \to j}=1\). The outcome is
\(Y^V_{ij}=\texttt{visit\_view}_{i \to j}\), which equals one if the exposure
leads to a campaign detail-page view. Our baseline linear probability
specification is
\begin{equation}
\begin{split}
Y^V_{ij}
&=
\sum_g \theta^{V,\mathrm{same}}_g
    \left(\texttt{local}_{i \to j} \times \mathbb{1}\{j \in g\}\right) \\
&\quad +
\sum_g \theta^{V,\mathrm{disp}}_g
    \left(\texttt{local}_{i \to j} \times \texttt{display}_{j}
    \times \mathbb{1}\{j \in g\}\right)
    + u_i + c_j + \varepsilon_{ij},
\end{split}
\label{eq:view_stage}
\end{equation}
where \(u_i\) denotes user fixed effects and \(c_j\) denotes campaign fixed
effects. The variables \(\texttt{local}_{i \to j}\), \(\texttt{adj}_{i \to j}\),
and \(\texttt{display}_{j}\) are defined in \Cref{tab:notations_definitions}.
The group indicator \(\mathbb{1}\{j \in g\}\) is defined by the headquarters
region of the campaign firm. In \Cref{table:main_tma_results}, \(g\) indexes
the two campaign regions, TMA and non-TMA. In
\Cref{table:area_lpm_results}, \(g\) indexes the finer economic-area groups.

The coefficient \(\theta^{V,\mathrm{same}}_g\) measures the same-prefecture
difference in the probability of viewing a campaign detail page when the
campaign listing does not display geographic information. The coefficient
\(\theta^{V,\mathrm{disp}}_g\) measures the additional same-prefecture
difference when the listing displays geographic information. Thus, for
campaigns with no listing-stage geographic display, the same-prefecture
difference is \(\theta^{V,\mathrm{same}}_g\); for campaigns with such a
display, the same-prefecture difference is
\(\theta^{V,\mathrm{same}}_g+\theta^{V,\mathrm{disp}}_g\).

Because \(\texttt{display}_{j}\) is defined at the campaign level, its main
effect is absorbed by campaign fixed effects. The coefficient
\(\theta^{V,\mathrm{disp}}_g\) therefore does not measure whether campaigns
with geographic displays receive more views overall. It measures whether
same-prefecture users are differentially more likely to view the same campaign
when the listing contains geographic information. Because geographic display is
not randomly assigned, we do not interpret
\(\theta^{V,\mathrm{disp}}_g\) as the causal effect of adding a geographic cue
to a listing. The coefficient is used as descriptive evidence on differential
viewing patterns when geographic cues are visible at the listing stage.

\subsection{Investment-Margin Specification}

The investment-margin specification follows the campaign-level participation
margin described in \Cref{sec:institution}. During our sample period, the
ordinary retail investor cap applied at the issuer level rather than as an
annual platform-wide investment budget. Participation in one campaign therefore
did not mechanically reduce the regulatory capacity available for other
campaigns. Together with campaign-specific minimum investment amounts, this
makes positive net investment at the user-campaign level a natural
extensive-margin outcome.

The investment-margin sample consists of user-campaign observations with a
campaign detail-page view. The outcome is \(Y^I_{ij}=\texttt{inv}_{i \to j}\),
defined for observations with \(\texttt{view}_{i \to j}=1\). It equals one if
user \(i\) has a positive net investment in campaign \(j\) and the first
observed detail-page view occurs no later than the first positive net
investment. Our baseline linear probability specification is
\begin{equation}
\begin{split}
Y^I_{ij}
&=
\sum_g \theta^{I,\mathrm{adj}}_g
    \left(\texttt{adj}_{i \to j} \times \mathbb{1}\{j \in g\}\right) \\
&\quad +
\sum_g \theta^{I,\mathrm{same}}_g
    \left(\texttt{local}_{i \to j} \times \mathbb{1}\{j \in g\}\right)
    + u_i + c_j + \eta_{ij},
\end{split}
\label{eq:invest_stage}
\end{equation}
where \(u_i\) and \(c_j\) again denote user and campaign fixed effects. The
group indicator is again defined by the campaign firm's headquarters region.
The coefficient \(\theta^{I,\mathrm{same}}_g\) measures the
investment-probability difference for same-prefecture viewers, relative to
other viewers of the same campaign and conditional on user fixed effects. The
coefficient \(\theta^{I,\mathrm{adj}}_g\) measures the corresponding difference
for adjacent-prefecture viewers, excluding same-prefecture pairs. Because the
investment-margin sample conditions on campaign-page viewing, the estimates
measure the observed transition from campaign-page access to investment rather
than initial campaign discovery. Conditioning on a detail-page view does not
imply that users devote the same attention to the page or interpret the
disclosed information in the same way. The estimates should not be read as
implying that same-prefecture and other-prefecture viewers have identical
latent interest in the campaign, or as eliminating all attention- or
information-related channels.

Comparing \(\theta^{I,\mathrm{same}}_g\) with
\(\theta^{I,\mathrm{adj}}_g\) implements the adjacent-prefecture comparison
described above. A positive same-prefecture coefficient together with a small
adjacent-prefecture coefficient indicates that the investment-margin difference
is concentrated among same-prefecture viewers rather than extending to nearby
prefectures more broadly.

\subsection{Fixed Effects, Support, and Remaining Channels}

The two-way fixed effects determine the identifying variation. Campaign fixed
effects compare users within the same campaign and absorb campaign-level
attributes common to users, including issuer characteristics, offering terms,
timing, and baseline campaign appeal. User fixed effects compare campaigns
within the same user and absorb time-invariant user characteristics, such as
residence location, persistent investment capacity, and baseline propensities
to view or invest. They do not absorb time-varying user states, including
platform experience accumulated over time.

The coefficients are therefore identified from residual variation across
user-campaign pairs. In the viewing-margin specification, the comparison is
whether same-prefecture exposed users are more likely than other exposed users
of the same campaign to proceed to the detail page. In the investment-margin
specification, the comparison is whether same-prefecture viewers are more likely
than other viewers of the same campaign to invest. The institutional setting
also helps interpret the investment margin: because the retail cap is
issuer-level rather than platform-wide, the binary outcome is less likely to
reflect regulatory capacity exhausted by investments in other campaigns.
Because these estimates rely on within-user and within-campaign overlap,
\Cref{tab:fe_support_diagnostics} documents, in the appendix, the overlap
underlying the relevant geographic comparisons and the distribution of user
activity. The table shows that substantial user overlap remains in both margins
and that the samples are not dominated by the most active users.

The fixed effects do not absorb pair-specific channels. Prior familiarity with
the issuer, offline ties, regionally targeted promotion, or other
user-campaign-specific information may vary within the same user and the same
campaign. These limitations matter, but the ECF setting reduces the force of
broad public-information accounts: issuers are small unlisted firms, and
detailed campaign information is provided through a common platform page. The
estimates therefore leave open pair-specific local channels. We use the
estimates together with the
adjacent-prefecture and TMA/non-TMA comparisons to assess whether the pattern is
better described by discovery alone, smooth proximity, or same-prefecture
locality. \Cref{sec:robustness_checks} reports a
conservative stress test that excludes a broad group of likely connected
investors, while recognizing that such a screen cannot remove all prior ties.

\subsection{Estimation and Inference}

Our baseline specifications are linear probability models with user and
campaign fixed effects. We interpret these estimates as linear projections of
conditional transition probabilities, not as structural models of individual
choice probabilities. The coefficients are therefore directly interpretable as
percentage-point differences.

Because the LPM does not constrain fitted values to lie between zero and one,
we also estimate logit specifications with the same right-hand-side variables
and report average marginal effects as nonlinear checks. We use the LPM
estimates as the main magnitude estimates because they provide transparent
two-way fixed-effect comparisons, whereas fixed-effect logit specifications can
drop observations with no within-group outcome variation and the logit
coefficients are not directly interpretable as probability differences.

Standard errors are two-way clustered by user and campaign, allowing arbitrary
correlation among observations for the same user and among observations for the
same campaign. Unless otherwise noted, the results reported below use the
two-way fixed-effects specifications described in this section.

%% file: sec_05_results.tex
\section{Empirical Results}
\label{sec:results}

\newcommand{\sym}[1]{\rlap{#1}}
\providecommand{\estse}[2]{\makecell[tc]{#1\\{\scriptsize (#2)}}}

This section reports the two-way fixed-effects estimates described in
\Cref{sec:empirical_strategy}. We interpret the LPM estimates as the main
magnitude estimates because they are directly expressed as percentage-point
differences in the relevant transition probabilities.
Logit average marginal effects are reported as nonlinear checks. The results
are organized around the empirical comparisons in \Cref{sec:empirical_strategy}:
the viewing margin, the investment margin, the geographic contrasts that link
them, and robustness checks for the main interpretation.

\subsection{Main Regional Results}

\Cref{table:main_tma_results} summarizes the main estimates after separating
campaigns located in the Tokyo metropolitan area (TMA) from campaigns located
elsewhere. In interpreting the estimates, we distinguish statistical
significance from economic magnitude. The descriptive rates in
\Cref{tab:analysis_sample_summary} provide useful benchmarks for scale, even
though they are campaign-level averages rather than the exact fitted baseline
from the fixed-effects regressions.

The listing-to-view columns report estimates for the viewing margin. For TMA
campaigns, the same-prefecture difference is statistically significant but
small when the listing does not display geographic information: the LPM
estimate is 0.004, or 0.4 percentage points. The additional same-prefecture
difference when the listing displays geographic information is also small and
not statistically significant.

For non-TMA campaigns, the corresponding viewing-margin estimates are much
larger. The same-prefecture difference is 0.027, or 2.7 percentage points, when
the listing does not display geographic information. The additional
same-prefecture difference when the listing displays geographic information is
0.093, or 9.3 percentage points. Thus, same-prefecture users are more likely to
inspect non-TMA campaigns after listing-page exposure, with a larger
same-prefecture viewing difference among campaigns whose listings display
geographic cues. The logit average marginal effects show the same
qualitative pattern.

The view-to-investment columns report the investment-margin estimates, which are
central to the interpretation. Among users who have viewed the campaign detail
page, the same-prefecture difference in the probability of positive net
investment is 0.006, or 0.6 percentage points, for TMA campaigns and 0.067, or
6.7 percentage points, for non-TMA campaigns. The adjacent-prefecture
difference is close to zero for TMA campaigns and 0.009, or 0.9 percentage
points, for non-TMA campaigns. Relative to the descriptive non-TMA
view-to-investment rate of about 10 percent for other-prefecture viewers, the
6.7 percentage-point same-prefecture difference is large, while the
adjacent-prefecture difference is modest. The central result is the non-TMA
investment-margin premium: same-prefecture viewers remain much more likely to
invest after campaign-page access, while the adjacent-prefecture difference is
modest. This is the key empirical pattern in the paper: the local premium is
not exhausted at the inspection margin, does not extend comparably to adjacent
prefectures, and is concentrated outside the metropolitan core. The nonlinear
checks do not alter these substantive conclusions.

\begin{table}[H]
\centering
\begin{threeparttable}
\caption{Main Two-Way Fixed-Effects Estimates by Campaign Region}
\label{table:main_tma_results}
\footnotesize
\setlength{\tabcolsep}{10pt}
\begin{tabular}{@{}llcccc@{}}
    \toprule
    & & \multicolumn{2}{c}{\textbf{Listing-to-view}} &
    \multicolumn{2}{c}{\textbf{View-to-investment}} \\
    \cmidrule(lr){3-4}\cmidrule(l){5-6}
    \textbf{Campaign region} & \textbf{Model} &
    \makecell[c]{Same\\pref.} &
    \makecell[c]{Same pref.\\\(\times\) display} &
    \makecell[c]{Adj.\\pref.} &
    \makecell[c]{Same\\pref.} \\
    \midrule
    TMA & LPM &
    \estse{0.004\sym{**}}{0.002} &
    \estse{0.006}{0.006} &
    \estse{0.000}{0.002} &
    \estse{0.006\sym{*}}{0.003} \\
    & Logit AME &
    0.005 & 0.007 & -0.000 & 0.006 \\
    \addlinespace
    Non-TMA & LPM &
    \estse{0.027\sym{***}}{0.005} &
    \estse{0.093\sym{**}}{0.030} &
    \estse{0.009\sym{**}}{0.003} &
    \estse{0.067\sym{***}}{0.007} \\
    & Logit AME &
    0.032 & 0.109 & 0.010 & 0.092 \\
    \bottomrule
\end{tabular}
\begin{tablenotes}[flushleft]
\footnotesize
\item[] 
\textbf{Notes:} LPM entries report coefficients from linear probability
models with user and campaign fixed effects. Coefficients are probability
differences; for example, 0.004 corresponds to 0.4 percentage points. Standard
errors clustered two ways by user and campaign are in parentheses. Logit AME
entries report average marginal effects from fixed-effect logit specifications
and are included as nonlinear checks. Stars refer to LPM entries.
*\(p<0.05\),
**\(p<0.01\), ***\(p<0.001\).
\end{tablenotes}
\end{threeparttable}
\end{table}

\subsection{Economic-Area Heterogeneity}

\Cref{table:area_lpm_results} reports a supplementary heterogeneity check that
disaggregates the non-TMA category into economic areas. The purpose is not to
replace the main TMA/non-TMA contrast, but to assess whether the main pattern is
driven by a single non-TMA area or appears more broadly across regional groups.
The estimates should be read with care because some display-support cells are
small, and the main regional contrast in \Cref{table:main_tma_results} remains
the primary summary.

The viewing-margin estimates show that \(\theta^{V,\mathrm{disp}}_g\) is small
for TMA campaigns but positive and large in several non-TMA economic areas. In
particular, the display interaction is large in Aichi, Chubu excluding Aichi,
Kansai/Chugoku outside Keihanshin, and Tohoku/Northern Kanto. Some areas have imprecise or
unstable estimates because only a few displayed campaigns contribute to the
interaction term. Fukuoka, for example, has only one displayed campaign and a
negative display interaction.

By contrast, all reported non-TMA same-prefecture investment estimates are
positive. The corresponding adjacent-prefecture estimates are generally smaller
and often not statistically significant, with Keihanshin as the clearest
exception. These estimates provide supporting evidence that the non-TMA
investment-margin premium is not confined to a single regional group, although
the adjacent estimate for Keihanshin indicates that regional proximity may
still matter in some areas.

\begin{table}[htbp]
\centering
\begin{threeparttable}
\caption{Economic-Area Heterogeneity in LPM Estimates}
\label{table:area_lpm_results}
\footnotesize
\setlength{\tabcolsep}{9pt}
\begin{tabular}{@{}lcccc@{}}
    \toprule
    & \multicolumn{2}{c}{\textbf{Listing-to-view}} &
    \multicolumn{2}{c}{\textbf{View-to-investment}} \\
    \cmidrule(lr){2-3}\cmidrule(l){4-5}
    \textbf{Campaign area} &
    \makecell[c]{Same\\pref.} &
    \makecell[c]{Same pref.\\\(\times\) display} &
    \makecell[c]{Adj.\\pref.} &
    \makecell[c]{Same\\pref.} \\
    \midrule
    TMA &
    \estse{0.004\sym{**}}{0.002} &
    \estse{0.006}{0.006} &
    \estse{0.000}{0.002} &
    \estse{0.006\sym{*}}{0.003} \\
    \addlinespace[1.5pt]
    Aichi &
    \estse{0.001}{0.009} &
    \estse{0.103\sym{***}}{0.011} &
    \estse{0.004}{0.011} &
    \estse{0.039\sym{***}}{0.012} \\
    \addlinespace[1.5pt]
    Chubu, excl. Aichi &
    \estse{0.054\sym{*}}{0.027} &
    \estse{0.225\sym{***}}{0.035} &
    \estse{0.004}{0.004} &
    \estse{0.125\sym{***}}{0.031} \\
    \addlinespace[1.5pt]
    Fukuoka &
    \estse{0.055\sym{***}}{0.016} &
    \estse{-0.119\sym{***}}{0.031} &
    \estse{-0.002}{0.020} &
    \estse{0.097\sym{***}}{0.022} \\
    \addlinespace[1.5pt]
    Hokkaido &
    \estse{0.096\sym{*}}{0.040} &
    \estse{0.070}{0.053} &
    -- &
    \estse{0.074\sym{**}}{0.028} \\
    \addlinespace[1.5pt]
    Keihanshin &
    \estse{0.015\sym{*}}{0.006} &
    \estse{0.044}{0.033} &
    \estse{0.014\sym{**}}{0.005} &
    \estse{0.042\sym{***}}{0.007} \\
    \addlinespace[1.5pt]
    Kansai/Chugoku, excl. Keihanshin &
    \estse{0.015}{0.033} &
    \estse{0.361\sym{***}}{0.042} &
    \estse{0.024}{0.017} &
    \estse{0.122\sym{**}}{0.037} \\
    \addlinespace[1.5pt]
    Kyushu/Shikoku, excl. Fukuoka &
    \estse{0.182\sym{***}}{0.034} &
    \estse{0.071}{0.082} &
    \estse{0.001}{0.009} &
    \estse{0.228\sym{***}}{0.041} \\
    \addlinespace[1.5pt]
    Tohoku/Northern Kanto &
    \estse{0.042}{0.030} &
    \estse{0.143\sym{***}}{0.039} &
    \estse{0.006}{0.007} &
    \estse{0.119\sym{***}}{0.036} \\
    \bottomrule
\end{tabular}
\begin{tablenotes}[flushleft]
\footnotesize
\item[] 
\textbf{Notes:} Entries report LPM coefficients from two-way
fixed-effects specifications; standard errors clustered two ways by user and
campaign are in parentheses. Coefficients are probability differences.
Hokkaido has no adjacent-prefecture observations under our adjacency definition.
*\(p<0.05\), **\(p<0.01\), ***\(p<0.001\).
\end{tablenotes}
\end{threeparttable}
\end{table}

\subsection{Economic Interpretation}

The results combine into a more specific pattern than a general geography
effect. The viewing-margin estimates show same-prefecture differences in
initial inspection for non-TMA campaigns, especially among listings with
visible geographic cues. Because the display evidence is descriptive, the main
interpretation comes from the same-prefecture premium in the
view-to-investment margin after observed campaign-page access.

That investment-margin premium is difficult to reconcile with a discovery-only
account, because it remains after users open the campaign page. It is also less
consistent with a smooth prefecture-level proximity account, because the
same-prefecture premium is substantially larger than the adjacent-prefecture
difference. Together with the small TMA premium, the pattern is better
described as same-prefecture locality: a same-prefecture investment pattern
that remains after campaign-page access, is larger than the
adjacent-prefecture difference, and is concentrated outside the metropolitan
core.

The estimates do not identify whether the remaining premium reflects local
information, prior ties, issuer outreach, or home-prefecture support motives,
including local attachment. The contribution is instead to show that a
substantial non-TMA local premium remains at the post-access investment margin
and is concentrated at the same-prefecture rather than adjacent-prefecture
margin.

\subsection{Robustness Checks}
\label{sec:robustness_checks}

Two additional checks address specific concerns about the interpretation. The
first concerns the viewing-margin display result. If same-prefecture users
appear more responsive when geography is visible on the listing page, the same
pattern might also appear for adjacent-prefecture users. We therefore estimate
an expanded viewing-margin specification that includes separate display
interactions for adjacent-prefecture users. In the pooled model, the
adjacent-display interaction is positive but much smaller than the
same-prefecture display interaction. In the TMA/non-TMA specification, the
non-TMA same-prefecture display interaction remains large, while the
corresponding non-TMA adjacent-display interaction is small and statistically
insignificant. The viewing-margin display result is therefore not simply a
nearby-prefecture display response.

The second check addresses pre-existing ties. Because issuers may contact
acquaintances and business networks, some same-prefecture investment could come
from users who already had a connection to the issuer. As a stress test, we
exclude likely connected investors, defined as users who invested within 21
days of registration, invested in only one campaign, and viewed fewer than five
campaign detail pages in total. This screen is intentionally broad: it may
remove some ordinary users, but it targets users for whom direct issuer contact
is particularly plausible.

\Cref{table:no_connected_results} reports the investment-margin estimates after
this exclusion. The non-TMA same-prefecture estimate falls from 0.067 to 0.051,
or from 6.7 to 5.1 percentage points, but remains statistically significant and
economically meaningful relative to the descriptive non-TMA other-prefecture
investment rate of about 10 percent. The adjacent-prefecture estimate also
falls slightly. This exercise does not remove all possible pair-specific local
channels, but the result cannot be attributed solely to the most direct form of
issuer-network mobilization captured by this screen.

\begin{table}[htbp]
\centering
\begin{threeparttable}
\caption{Robustness: View-to-Investment Estimates Excluding Likely Connected Investors}
\label{table:no_connected_results}
\footnotesize
\setlength{\tabcolsep}{12pt}
\begin{tabular}{@{}llcc@{}}
    \toprule
    & & \multicolumn{2}{c}{\textbf{View-to-investment}} \\
    \cmidrule(l){3-4}
    \textbf{Campaign region} & \textbf{Model} &
    \makecell[c]{Adj.\\pref.} &
    \makecell[c]{Same\\pref.} \\
    \midrule
    TMA & LPM &
    \estse{-0.000}{0.002} &
    \estse{0.004}{0.003} \\
    & Logit AME &
    -0.001 & 0.004 \\
    \addlinespace
    Non-TMA & LPM &
    \estse{0.007\sym{*}}{0.003} &
    \estse{0.051\sym{***}}{0.006} \\
    & Logit AME &
    0.008 & 0.069 \\
    \bottomrule
\end{tabular}
\begin{tablenotes}[flushleft]
\footnotesize
\item[] 
\textbf{Notes:} Entries are view-to-investment estimates after excluding
likely connected investors. LPM entries report coefficients from linear
probability models with user and campaign fixed effects; standard errors
clustered two ways by user and campaign are in parentheses. Logit AME entries
report average marginal effects from fixed-effect logit specifications and are
included as nonlinear checks. Stars refer to LPM entries.
*\(p<0.05\), **\(p<0.01\),
***\(p<0.001\).
\end{tablenotes}
\end{threeparttable}
\end{table}

%% file: conclusion.tex
\section{Concluding Remarks}

Digital platforms can make distant investment opportunities accessible, but
access does not make capital allocation geographically neutral. On Fundinno,
users reach campaign pages through a common online interface and can access
detailed campaign information before investing. The central question is
therefore not simply whether online platforms reduce search frictions, but how
remaining geographic concentration should be interpreted once exposure,
campaign-page access, and investment can be observed separately.

The main result is a same-prefecture premium after campaign-page access for
campaigns outside the Tokyo metropolitan area. For non-TMA campaigns, viewers
from the same prefecture are 6.7 percentage points more likely to make a
positive net investment, relative to an other-prefecture investment rate of
roughly 10 percent. The adjacent-prefecture difference is only 0.9 percentage
points, and the corresponding premium is small for TMA campaigns. Together,
these estimates show that the main local premium remains after observed
campaign-page access and is much larger than the adjacent-prefecture
difference.

This pattern is more specific than a general geography effect. It is not simply
that geographically nearby users invest more: the premium is concentrated among
same-prefecture users, remains after campaign-page access, and is muted in the
metropolitan core. We therefore interpret the evidence as pointing to
same-prefecture locality, while recognizing that the data do not identify a
single motive. Local information, prior ties, issuer outreach, and
home-prefecture support motives, including local attachment, remain possible
channels.

The broader implication is that online access and geographic neutrality are
distinct. A common digital interface can widen the set of investment
opportunities available to users while leaving investment demand geographically
bounded. For regional issuers, especially outside the
metropolitan core, same-prefecture users may remain an important source of
investment demand after campaign-page access even on a national online
platform. From a regional-development perspective, online ECF may be useful not
because it eliminates local dependence, but because it allows regional issuers
to combine broader online reach with home-prefecture demand. In this sense,
online equity finance can broaden access without eliminating the local
structure of investment demand.

%% file: appendix.tex
\newpage
\section*{Appendix}
\providecommand{\sym}[1]{\rlap{#1}}
\providecommand{\estse}[2]{\makecell[tc]{#1\\{\scriptsize (#2)}}}

\subsection*{Appendix A. Regional Classification and Adjacency}

\footnotesize
\renewcommand{\arraystretch}{0.95}
\begin{longtable}{@{}rlll@{}}
\caption{Regional Classification and Adjacency of Japanese Prefectures}
\label{table:adjacency} \\
\toprule
 & Prefecture & Area & Adjacent prefectures \\
\midrule
\endfirsthead
\caption[]{Regional Classification and Adjacency of Japanese Prefectures} \\
\toprule
 & Prefecture & Area & Adjacent prefectures \\
\midrule
\endhead
\midrule
\multicolumn{4}{r}{Continued on next page} \\
\midrule
\endfoot
\bottomrule
\endlastfoot
1 & Hokkaido & HKD & None \\
2 & Aomori & TOH-NKT & Iwate, Akita \\
3 & Iwate & TOH-NKT & Aomori, Akita, Miyagi \\
4 & Miyagi & TOH-NKT & Iwate, Akita, Yamagata, Fukushima \\
5 & Akita & TOH-NKT & Aomori, Iwate, Miyagi, Yamagata \\
6 & Yamagata & TOH-NKT & Akita, Miyagi, Fukushima, Niigata \\
7 & Fukushima & TOH-NKT & Miyagi, Yamagata, Niigata, Gunma, Tochigi, Ibaraki \\
8 & Ibaraki & TOH-NKT & Fukushima, Tochigi, Saitama, Chiba \\
9 & Tochigi & TOH-NKT & Fukushima, Ibaraki, Gunma, Saitama \\
10 & Gunma & TOH-NKT & Fukushima, Tochigi, Saitama, Nagano, Niigata \\
11 & Saitama & TMA & Gunma, Tochigi, Ibaraki, Chiba, Tokyo, Yamanashi, Nagano \\
12 & Chiba & TMA & Ibaraki, Saitama, Tokyo \\
13 & Tokyo & TMA & Saitama, Chiba, Kanagawa, Yamanashi \\
14 & Kanagawa & TMA & Tokyo, Yamanashi, Shizuoka \\
15 & Niigata & CHU & Yamagata, Fukushima, Gunma, Nagano, Toyama \\
16 & Toyama & CHU & Niigata, Nagano, Gifu, Ishikawa \\
17 & Ishikawa & CHU & Toyama, Gifu, Fukui \\
18 & Fukui & CHU & Ishikawa, Gifu, Shiga, Kyoto \\
19 & Yamanashi & CHU & Saitama, Tokyo, Kanagawa, Shizuoka, Nagano \\
20 & Nagano & CHU & Niigata, Gunma, Saitama, Yamanashi, Shizuoka, Aichi, Gifu, Toyama \\
21 & Gifu & CHU & Toyama, Ishikawa, Nagano, Aichi, Mie, Shiga, Fukui \\
22 & Shizuoka & CHU & Kanagawa, Yamanashi, Nagano, Aichi \\
23 & Aichi & AIC & Shizuoka, Nagano, Gifu, Mie \\
24 & Mie & KNS-CHG & Aichi, Gifu, Shiga, Kyoto, Nara, Wakayama \\
25 & Shiga & KHS & Fukui, Gifu, Mie, Kyoto \\
26 & Kyoto & KHS & Fukui, Shiga, Mie, Nara, Osaka, Hyogo \\
27 & Osaka & KHS & Kyoto, Nara, Hyogo, Wakayama \\
28 & Hyogo & KHS & Kyoto, Osaka, Okayama, Tottori \\
29 & Nara & KNS-CHG & Kyoto, Osaka, Wakayama, Mie \\
30 & Wakayama & KNS-CHG & Osaka, Nara, Mie \\
31 & Tottori & KNS-CHG & Hyogo, Okayama, Hiroshima, Shimane \\
32 & Shimane & KNS-CHG & Tottori, Okayama, Hiroshima, Yamaguchi \\
33 & Okayama & KNS-CHG & Hyogo, Tottori, Shimane, Hiroshima, Kagawa \\
34 & Hiroshima & KNS-CHG & Tottori, Okayama, Yamaguchi, Ehime, Shimane \\
35 & Yamaguchi & KNS-CHG & Shimane, Hiroshima, Fukuoka \\
36 & Tokushima & KYS & Kagawa, Ehime, Kochi \\
37 & Kagawa & KYS & Tokushima, Ehime, Okayama \\
38 & Ehime & KYS & Kagawa, Tokushima, Kochi, Hiroshima \\
39 & Kochi & KYS & Tokushima, Ehime \\
40 & Fukuoka & FUK & Yamaguchi, Oita, Kumamoto, Saga \\
41 & Saga & KYS & Fukuoka, Nagasaki \\
42 & Nagasaki & KYS & Saga \\
43 & Kumamoto & KYS & Fukuoka, Oita, Miyazaki, Kagoshima \\
44 & Oita & KYS & Fukuoka, Kumamoto, Miyazaki \\
45 & Miyazaki & KYS & Oita, Kumamoto, Kagoshima \\
46 & Kagoshima & KYS & Kumamoto, Miyazaki \\
47 & Okinawa & KYS & None \\
\end{longtable}

\noindent\parbox{0.95\linewidth}{%
\footnotesize
\textit{Notes:} Area abbreviations are HKD = Hokkaido, TOH-NKT = Tohoku and
Northern Kanto, TMA = Tokyo Metropolitan Area, CHU = Chubu excluding Aichi,
AIC = Aichi, KHS = Keihanshin, KNS-CHG = Kansai and Chugoku outside
Keihanshin, FUK = Fukuoka, and KYS = Kyushu and Shikoku excluding Fukuoka.
Hokkaido is coded as having no adjacent prefecture.}

\renewcommand{\arraystretch}{1}
\normalsize

\subsection*{Appendix B. Economic-Area Sample Support}

\begin{table}[H]
\centering
\begin{threeparttable}
\caption{Geographic Distribution by Economic Area}
\label{tab:geo_area_distribution}
\footnotesize
\newcommand{\countpct}[2]{\makebox[3.3em][r]{#1}\hspace{0.2em}\makebox[3.9em][r]{(#2\%)}}
\begin{tabular}{@{}lccc@{}}
    \toprule
    Economic area & Campaigns & Display listings & Eligible users \\
    \midrule
    TMA & \countpct{383}{72.7} & \countpct{9}{2.3} & \countpct{24,755}{50.9} \\
    Aichi & \countpct{15}{2.8} & \countpct{1}{6.7} & \countpct{2,851}{5.9} \\
    Chubu, excl. Aichi & \countpct{16}{3.0} & \countpct{4}{25.0} & \countpct{3,323}{6.8} \\
    Fukuoka & \countpct{17}{3.2} & \countpct{1}{5.9} & \countpct{1,705}{3.5} \\
    Hokkaido & \countpct{8}{1.5} & \countpct{4}{50.0} & \countpct{1,116}{2.3} \\
    Keihanshin & \countpct{45}{8.5} & \countpct{4}{8.9} & \countpct{6,735}{13.8} \\
    Kansai/Chugoku, excl. Keihanshin & \countpct{6}{1.1} & \countpct{2}{33.3} & \countpct{2,744}{5.6} \\
    Kyushu/Shikoku, excl. Fukuoka & \countpct{24}{4.6} & \countpct{6}{25.0} & \countpct{2,405}{4.9} \\
    Tohoku/Northern Kanto & \countpct{13}{2.5} & \countpct{2}{15.4} & \countpct{3,032}{6.2} \\
    \midrule
    Total & \countpct{527}{100.0} & \countpct{33}{6.3} & \countpct{48,666}{100.0} \\
    \bottomrule
\end{tabular}
\begin{tablenotes}[flushleft]
\footnotesize
\item[] 
\textbf{Notes:} Campaigns are classified by the headquarters area of the
issuer. Eligible users are classified by residence area. Display listings are
campaigns whose listing text displays a prefecture or city name. Percentages
for campaigns and eligible users are shares of the corresponding column total.
Percentages for display listings are shares of campaigns within the same
economic area. Three eligible users have missing residence prefecture and are
excluded from the user area shares.
\end{tablenotes}
\end{threeparttable}
\end{table}

\subsection*{Appendix C. Support for Within-User and Within-Campaign Comparisons}

\Cref{tab:fe_support_diagnostics} documents the overlap underlying the
within-user and within-campaign comparisons used in the fixed-effect
specifications. In the
listing-to-view sample, 46.9 percent of users have both same-prefecture and
other-prefecture observations; in the view-to-investment sample, the
corresponding share is 35.4 percent. The most active one percent of users
account for 13.1 percent and 18.3 percent of observations in the two samples,
respectively, indicating that the estimates are not driven solely by a small set
of highly active users.

\begin{table}[H]
\centering
\begin{threeparttable}
\caption{Support for Within-User and Within-Campaign Comparisons}
\label{tab:fe_support_diagnostics}
\footnotesize
\setlength{\tabcolsep}{5pt}
\begin{tabular}{@{}lrrrrrrr@{}}
    \toprule
    & & & &
    \multicolumn{3}{c}{\textbf{Obs. per user}} & \\
    \cmidrule(lr){5-7}
    \textbf{Margin} &
    \textbf{Obs.} &
    \textbf{Users} &
    \makecell[c]{\textbf{User}\\\textbf{overlap}} &
    \textbf{Median} &
    \textbf{P90} &
    \textbf{P99} &
    \makecell[c]{\textbf{Top 1\%}\\\textbf{share}} \\
    \midrule
    Listing-to-view &
    1,179,437 &
    42,913 &
    \makecell[r]{20,145\\(46.9\%)} &
    9 & 71 & 284 & 13.1\% \\
    View-to-investment &
    574,213 &
    35,527 &
    \makecell[r]{12,565\\(35.4\%)} &
    4 & 37 & 206 & 18.3\% \\
    \bottomrule
\end{tabular}
\begin{tablenotes}[flushleft]
\footnotesize
\item[] 
\textbf{Notes:} The table summarizes support in the two model samples.
The user overlap column counts users with both same-prefecture and
other-prefecture observations in the relevant margin. All 527 campaigns have
both same-prefecture and other-prefecture observations in both margins, and 516
campaigns have at least one adjacent-prefecture observation. User-activity
percentiles are calculated from observations per user in the corresponding
sample. The top 1\% share is the fraction of sample observations accounted for
by the most active one percent of users.
\end{tablenotes}
\end{threeparttable}
\end{table}

\subsection*{Appendix D. Viewing-Stage Adjacent-Prefecture Check}

\begin{table}[H]
\centering
\begin{threeparttable}
\caption{Viewing-Stage Estimates with Adjacent-Prefecture Terms}
\label{tab:view_adjacent_display_check}
\footnotesize
\setlength{\tabcolsep}{6pt}
\begin{tabular}{@{}llcccc@{}}
    \toprule
    & &
    \makecell[c]{Same\\pref.} &
    \makecell[c]{Same pref.\\\(\times\) display} &
    \makecell[c]{Adj.\\pref.} &
    \makecell[c]{Adj. pref.\\\(\times\) display} \\
    \midrule
    All campaigns & LPM &
    \estse{0.009\sym{***}}{0.002} &
    \estse{0.038\sym{**}}{0.013} &
    \estse{0.002}{0.001} &
    \estse{0.011\sym{***}}{0.003} \\
    & Logit AME & 0.010 & 0.044 & 0.003 & 0.014 \\
    \addlinespace[0.15em]
    TMA & LPM &
    \estse{0.004\sym{*}}{0.002} &
    \estse{0.009}{0.007} &
    \estse{-0.001}{0.002} &
    \estse{0.010\sym{*}}{0.004} \\
    & Logit AME & 0.005 & 0.011 & -0.001 & 0.012 \\
    \addlinespace[0.15em]
    Non-TMA & LPM &
    \estse{0.028\sym{***}}{0.005} &
    \estse{0.093\sym{**}}{0.030} &
    \estse{0.004}{0.003} &
    \estse{0.006}{0.006} \\
    & Logit AME & 0.033 & 0.109 & 0.004 & 0.007 \\
    \bottomrule
\end{tabular}
\begin{tablenotes}[flushleft]
\footnotesize
\item[] 
\textbf{Notes:} The dependent variable equals one if a qualifying
listing-page exposure leads to a campaign detail-page view. LPM entries are
two-way fixed-effect estimates; standard errors clustered by user and campaign
are in parentheses. Logit AME entries are nonlinear checks. Stars refer to LPM
entries. The LPM sample contains 1,176,484 observations and the fixed-effect
logit sample contains 1,095,913 observations. *\(p<0.05\), **\(p<0.01\),
***\(p<0.001\).
\end{tablenotes}
\end{threeparttable}
\end{table}

\subsection*{Appendix E. Nonlinear Robustness Checks}

\begin{table}[H]
\centering
\begin{threeparttable}
\caption{Economic-Area Fixed-Effect Logit Estimates}
\label{tab:area_logit_ame}
\footnotesize
\renewcommand{\arraystretch}{0.95}
\setlength{\tabcolsep}{6pt}
\begin{tabular}{@{}llcccc@{}}
    \toprule
    & & \multicolumn{2}{c}{\textbf{Listing-to-view}} &
    \multicolumn{2}{c}{\textbf{View-to-investment}} \\
    \cmidrule(lr){3-4}\cmidrule(l){5-6}
    \textbf{Campaign area} & \textbf{Estimate} &
    \makecell[c]{Same\\pref.} &
    \makecell[c]{Same pref.\\\(\times\) display} &
    \makecell[c]{Adj.\\pref.} &
    \makecell[c]{Same\\pref.} \\
    \midrule
    TMA & Logit &
    \estse{0.039\sym{**}}{0.014} &
    \estse{0.049}{0.056} &
    \estse{-0.003}{0.027} &
    \estse{0.061}{0.034} \\
    & AME & 0.005 & 0.007 & -0.000 & 0.006 \\
    \addlinespace[2pt]
    Aichi & Logit &
    \estse{0.012}{0.069} &
    \estse{0.698\sym{***}}{0.086} &
    \estse{0.046}{0.108} &
    \estse{0.392\sym{***}}{0.080} \\
    & AME & 0.002 & 0.118 & 0.005 & 0.049 \\
    \addlinespace[2pt]
    Chubu, excluding Aichi & Logit &
    \estse{0.453\sym{*}}{0.189} &
    \estse{1.633\sym{***}}{0.246} &
    \estse{0.086}{0.056} &
    \estse{1.461\sym{***}}{0.280} \\
    & AME & 0.068 & 0.290 & 0.007 & 0.189 \\
    \addlinespace[2pt]
    Fukuoka & Logit &
    \estse{0.382\sym{***}}{0.103} &
    \estse{-0.792\sym{***}}{0.239} &
    \estse{0.017}{0.233} &
    \estse{1.013\sym{***}}{0.182} \\
    & AME & 0.061 & -0.133 & 0.002 & 0.140 \\
    \addlinespace[2pt]
    Hokkaido & Logit &
    \estse{0.745\sym{*}}{0.296} &
    \estse{0.573}{0.355} &
    -- &
    \estse{0.919\sym{*}}{0.392} \\
    & AME & 0.115 & 0.093 & -- & 0.098 \\
    \addlinespace[2pt]
    Keihanshin & Logit &
    \estse{0.144\sym{**}}{0.044} &
    \estse{0.300}{0.215} &
    \estse{0.170\sym{**}}{0.065} &
    \estse{0.517\sym{***}}{0.082} \\
    & AME & 0.020 & 0.045 & 0.016 & 0.057 \\
    \addlinespace[2pt]
    Kansai/Chugoku, excl. Keihanshin & Logit &
    \estse{0.148}{0.250} &
    \estse{2.143\sym{***}}{0.316} &
    \estse{0.476\sym{*}}{0.191} &
    \estse{1.832\sym{***}}{0.306} \\
    & AME & 0.021 & 0.437 & 0.030 & 0.208 \\
    \addlinespace[2pt]
    Kyushu/Shikoku, excluding Fukuoka & Logit &
    \estse{1.201\sym{***}}{0.190} &
    \estse{0.540}{0.396} &
    \estse{-0.045}{0.154} &
    \estse{1.933\sym{***}}{0.272} \\
    & AME & 0.217 & 0.103 & -0.003 & 0.341 \\
    \addlinespace[2pt]
    Tohoku/Northern Kanto & Logit &
    \estse{0.366}{0.217} &
    \estse{1.186\sym{***}}{0.266} &
    \estse{0.021}{0.176} &
    \estse{1.618\sym{***}}{0.302} \\
    & AME & 0.053 & 0.217 & 0.001 & 0.193 \\
    \bottomrule
\end{tabular}
\begin{tablenotes}[flushleft]
\scriptsize
\item[] 
\textbf{Notes:} Logit rows report fixed-effect logit coefficients;
standard errors clustered two ways by user and campaign are in parentheses.
AME rows report average marginal effects and are probability differences; for
example, 0.005 corresponds to 0.5 percentage points. Stars refer to the logit
coefficients. Effective logit samples contain 1,095,913 observations for
listing-to-view and 498,230 observations for view-to-investment. Hokkaido has no
adjacent-prefecture observations under our adjacency definition. *\(p<0.05\),
**\(p<0.01\), ***\(p<0.001\).
\end{tablenotes}
\end{threeparttable}
\end{table}
\renewcommand{\arraystretch}{1}

%% file: ecf.bib
@incollection{Kato2021-yr,
  author    = {Kato, Takahito},
  title     = {The Legal Regulation of Crowdfunding in Japan},
  booktitle = {Legal Aspects of Crowdfunding},
  publisher = {Springer International Publishing},
  address   = {Cham},
  year      = {2021},
  pages     = {343--362},
  language  = {en}
}

@article{Nose2023-bk,
  author    = {Nose, Yoshiaki and Hosomi, Chie},
  title     = {What Makes Equity Crowdfunding Successful in Japan? Testing the Signaling and Lack of Financial Literacy Hypotheses},
  journal   = {Journal of Entrepreneurship Management and Innovation},
  publisher = {Fundacja Upowszechniajaca Wiedze i Nauke Cognitione},
  year      = {2023},
  volume    = {19},
  number    = {4},
  pages     = {146--183}
}

@article{Honjo2023-tl,
  author  = {Honjo, Yuji and Kurihara, Koki},
  title   = {Target for Campaign Success: An Empirical Analysis of Equity Crowdfunding in Japan},
  journal = {The Journal of Technology Transfer},
  year    = {2024},
  month   = jun
}

@article{Inaba2023-cv,
  author    = {Inaba, Kei-Ichiro and Komatsu, Masami and Miyakawa, Daisuke},
  title     = {A First Foray into the Phenomenon of Equity Crowdfunding in Japan: What Do Private Fundraising Records Tell?},
  journal   = {SSRN Electronic Journal},
  publisher = {Elsevier},
  year      = {2023},
  month     = jul,
  language  = {en}
}

@misc{FUNDINNO2022SpecifiedInvestor,
  author       = {{FUNDINNO, Inc.}},
  title        = {FUNDINNO Tokutei Toshika Seido no Donyu wo Kaishi},
  year         = {2022},
  howpublished = {\url{https://prtimes.jp/main/html/rd/p/000000171.000021941.html}},
  note         = {Accessed May 6, 2026}
}

@misc{JSDA2026ECFOverview,
  author       = {{Japan Securities Dealers Association}},
  title        = {Kabushiki Toshi-gata Crowdfunding: Seido no Gaiyo},
  year         = {2026},
  howpublished = {\url{https://market.jsda.or.jp/shijyo/kabucrowdfunding/seido/gaiyou/index.html}},
  note         = {Accessed May 6, 2026}
}

@misc{JSDA2026ECFStatistics,
  author       = {{Japan Securities Dealers Association}},
  title        = {Kabushiki Toshi-gata Crowdfunding no Tokei Joho / Toriatsukai Jokyo},
  year         = {2026},
  howpublished = {\url{https://www.jsda.or.jp/shiryoshitsu/toukei/kabucrowdfunding/index.html}},
  note         = {Excel statistics file accessed May 6, 2026}
}

@article{Agrawal2015-fr,
  author    = {Agrawal, Ajay and Catalini, Christian and Goldfarb, Avi},
  title     = {Crowdfunding: Geography, Social Networks, and the Timing of Investment Decisions},
  journal   = {Journal of Economics \& Management Strategy},
  publisher = {Wiley},
  year      = {2015},
  volume    = {24},
  number    = {2},
  pages     = {253--274},
  month     = may,
  language  = {en}
}

@article{Lin2016-kw,
  author    = {Lin, Mingfeng and Viswanathan, Siva},
  title     = {Home Bias in Online Investments: An Empirical Study of an Online Crowdfunding Market},
  journal   = {Management Science},
  publisher = {INFORMS},
  year      = {2016},
  volume    = {62},
  number    = {5},
  pages     = {1393--1414},
  month     = may,
  language  = {en}
}

@article{Guenther,
  author   = {Guenther, Christina and Johan, Sofia and Schweizer, Denis},
  title    = {Is the Crowd Sensitive to Distance? How Investment Decisions Differ by Investor Type},
  journal  = {Small Business Economics},
  year     = {2018},
  volume   = {50},
  number   = {2},
  pages    = {289--305},
  month    = feb,
  doi      = {10.1007/s11187-016-9834-6},
  url      = {https://ideas.repec.org/a/kap/sbusec/v50y2018i2d10.1007_s11187-016-9834-6.html}
}

@article{Bade2021-uq,
  author    = {Bade, Marco and Walther, Martin},
  title     = {Local Preferences and the Allocation of Attention in Equity-Based Crowdfunding},
  journal   = {Review of Managerial Science},
  publisher = {Springer Science and Business Media},
  year      = {2021},
  volume    = {15},
  number    = {8},
  pages     = {2501--2533},
  month     = nov,
  language  = {en}
}

@article{Hornuf2022-fs,
  author    = {Hornuf, Lars and Schmitt, Matthias and Stenzhorn, Eliza},
  title     = {The Local Bias in Equity Crowdfunding: Behavioral Anomaly or Rational Preference?},
  journal   = {Journal of Economics \& Management Strategy},
  publisher = {Wiley},
  year      = {2022},
  volume    = {31},
  number    = {3},
  pages     = {693--733},
  month     = mar,
  language  = {en}
}

@article{Cai2024-eg,
  author    = {Cai, Wanxiang and Polzin, Friedemann and Stam, Erik},
  title     = {Mitigating Local Bias in Equity Crowdfunding: A Financial Ecology Perspective},
  journal   = {Journal of Economic Geography},
  publisher = {Oxford University Press},
  year      = {2024},
  volume    = {24},
  number    = {4},
  pages     = {549--565},
  month     = jul,
  language  = {en}
}
